\documentclass[aps,bibnotes,twocolumn,showpacs, showkeys, superscriptaddress,preprintnumbers,amsmath,amssymb,lengthcheck,pra]{revtex4-1}

\usepackage{graphicx}
\usepackage{dcolumn}
\usepackage{bm}
\usepackage{color}

\begin{document}

\title{Akhmediev Breathers and Peregrine Solitary Waves in a Quadratic Medium}


\author{Fabio Baronio} 
\email{fabio.baronio@unibs.it}
\affiliation{INO CNR and Dipartimento di Ingegneria dell'Informazione, Universit\`a di Brescia, Via Branze 38, 25123 Brescia, Italy}

\begin{abstract}
We investigate the formation of optical localized nonlinear structures, evolving upon 
a non-zero background plane wave, in a dispersive quadratic medium.
We show the existence of quadratic Akhmediev breathers and Peregrine solitary waves, 
in the regime of cascading second-harmonic generation. 
This finding opens a novel path for the excitation of extreme rogue waves 
and for the description of modulation instability in quadratic nonlinear optics.

%
%
%
\end{abstract}

													

\maketitle

In dispersive optical media, with cubic Kerr nonlinearity, the nonlinear Schr\"odinger equation (NLSE) provides a central 
description of a variety of nonlinear localization effects \cite{ak97}. 
In fact, since the 1970s there has been wide and continued investigation into the properties of analytic soliton solutions of the NLSE \cite{za71}, 
both for their intrinsic scientific interest, as well as for their potential to provide new insights into important applications
such as optical propagation in nonlinear waveguides \cite{ag07}. For the case of a self-focussing nonlinearity, the most celebrated 
solution of this type is very likely the propagation-invariant hyperbolic secant bright soliton, but there is also an extensive 
literature studying various types of solitons on finite background consisting of a localized nonlinear structure evolving upon 
a non-zero background plane wave \cite{ma79,pe83,ak86}.

Solitons on finite background have recently attracted significant interest as their localization dynamics have been proposed 
as an important mechanism underlying the formation of the infamous extreme amplitude rogue waves on the surface of the 
ocean \cite{ka08,on16}. Much of this work has also been motivated by a parallel research effort using nonlinear optical fibre systems 
to implement controlled experiments studying NLSE dynamics and rogue waves in an optical context \cite{du09,ki10,ha11,on13,mu14,du14,rev16}. 
Many of the recent studies have focussed on the characteristics of the Akhmediev breather \cite{ak86}, a soliton on finite background 
solution which is excited from a weak periodic modulation and which is localized in the longitudinal dimension as it undergoes 
growth and decay. Experiments in optics have demonstrated important links to modulation instability (MI) \cite{du09,on13,mu14,du14}: 
of importance has been the realization that many properties of MI previously described only approximately 
(via numerical or truncated mode approaches) can in fact be described almost exactly using Akhmediev-breathers. 
Another significant application of the theory of Akhmediev breathers has been to design experiments generating the 
rational Peregrine soliton \cite{pe83}, an important and limiting case of a solitons on finite background solution that is
localised in both transverse and longitudinal dimensions \cite{ki10,ha11}.

In dispersive optical media, with quadratic nonlinearity, the existence of localized nonlinear structures evolving upon 
a non-zero background plane wave remained unexplored to date.

In this Letter, we investigate the formation of localized nonlinear structures, evolving upon 
a non-zero background plane wave, in optical systems described by second harmonic generation 
(SHG) equations \cite{boyd}. We show the existence of quadratic Akhmediev breathers and Peregrine solitary waves, 
in the regime of cascading SHG \cite{menyuk94}. 
To our knowledge, no studies have explicitly characterized nonlinear breather 
localization in any systems described by the SHG.
This finding opens a novel path for the excitation of 
extreme rogue waves, and for the description of MI \cite{tri95} in quadratic media. 

In transparent dispersive media, with quadratic nonlinearity,
the interaction of the fundamental frequency (FF)  and the second harmonic (SH) envelopes $u_{1,2}=u_{1,2}(\tau,\xi)$,
at frequency $\omega_0$ and $2\omega_0$ respectively,
obeys  coupled equations that, in dimensionless form, read as \cite{menyuk94,tri95}

\begin{eqnarray}\label{shg}
\nonumber & i u_{1 \xi}   -\frac{\beta_1}{2}u_{1 \tau \tau}+ u_2 u_1^* e^{-i\delta k \xi}=
0,\\
 &  i u_{2 \xi}  +i v u_{2 \tau} -\frac{\beta_2}{2}u_{2 \tau \tau}+ u_1^2 e^{i\delta k \xi}=0,
\end{eqnarray}
where $\xi \equiv z/z_d \equiv z |\beta_1|/t_0^2$ is the propagation distance in units of the
dispersion lenght $z_d=t_0^2/|\beta_1|$ associated with the FF dispersion $\beta''_1=d^2k/d \omega^2|_{\omega_0}$;
$\tau=(t-z/v_1)/t_0$ is the time in a reference frame traveling with the FF group velocity $v_1^{-1}=d k/d \omega|_{\omega_0}$ 
($t_0$ being an arbitrary time scale); $\beta_1=sign(\beta''_1)$, $\beta_2=-sign(\beta''_2)|\beta''_2/\beta''_1|$,
with $\beta''_2=d^2k/d \omega^2|_{2\omega_0}$; $\delta k= \Delta k z_d$ is the normalized wave-vector mismatch, with $\Delta k=k_2-2 k_1$, $k_1=k|_{\omega_0}$,
$k_2=k|_{2\omega_0}$;
$v \equiv z_d/z_w$ is the ratio between the dispersion and the walk-off length $z_w \equiv t_0/(v_2^{-1}-v_1^{-1})$, where 
$v_2^{-1}=d k/d \omega|_{2\omega_0}$ is the SH group velocity.
The fields $u_1=\chi z_d A_1$ and $u_2=\chi z_d A_2$, where $|A|_{1,2}^2$ measure directly the intensity 
(in watts per square meter) if we set the nonlinear coefficient $\chi=\omega_0[2/(c^3\epsilon_0 n^2_{\omega 0} n_{2 \omega 0})]^{1/2} d^{(2)}$
and $d^{(2)}$ is the effective susceptibility element (in meter per volt). 

We point out that Eqs. (\ref{shg}) also govern the interaction of beams focused in one transverse dimension,
making the substitution $t \rightarrow x $, $ \beta''_i \rightarrow 1/ 2 k_i $, and being  $v$ the birefingence
walk-off.
%

In the cascading regime, at large phase-mismatch, by using the SH asymptotic 
expansion and the method of repeated substitution, we arrive at a single FF evolution 
equation, at first order approximation \cite{menyuk94}:
\begin{equation}\label{CLL}
i \rho_ \xi -\frac{\beta_1}{2}\rho_{\tau \tau}+\sigma|\rho|^2 \rho=0,
\end{equation}
\noindent where, for sake of clarity, we have defined  $u_1=\rho$, $u_2 \simeq \rho^2 e^{i \delta k \xi} / \delta k$, 
$\sigma=1/\delta k$. 
%
We emphasize that the validity of the mapping of the FF wave evolution of Eqs. (\ref{shg}) into 
Eq. (\ref{CLL}) entails weak overall FF to SH conversion. 

Over the last decades, the cascading 
regime has been successfully exploited in the demonstration 
of quadratic spatial solitary waves \cite{schiek96},
soliton bouncing at nonlinear interfaces \cite{baro04}, 
temporal solitary waves \cite{ditra98}, 
steepening and spectral dynamics \cite{moses06,baro06}, 
temporal compression \cite{bache07,bache12},
and shock waves \cite{confo12}.  

In this Letter, we focus our attention on
i) anomalous dispersion ($\beta_1 = -1$) and effective self--focusing nonlinearities ($\sigma>0 $), 
ii) normal dispersion ($\beta_1 =1$) and effective self--defocusing nonlinearities ($\sigma <0 $), 
which both yield the integrable focusing NLSE.

%
%

We proceeded to demonstrate the existence of FF Akhmediev breathers of the SHG Eqs. (\ref{shg}),
which are predicted theoretically through the mapping with the NLSE Eq. (\ref{CLL}).
We consider the standard case in the experiments such that only a FF pulse is launched in
the quadratic medium. 

A form of the  Akhmediev breathers soliton solution of the NLSE Eq. (\ref{CLL}), under the condition $\beta_1 \sigma<0$,
can be expressed as \cite{ak86,ha11}:
\begin{equation}\label{AB}
\rho=\rho_0 \Big[ \frac{(1-4a)cosh(b \xi_n)+\sqrt{2a}cos(\Omega \tau_n)+i b sinh(b \xi_n)    }{\sqrt{2a}cos(\Omega \tau_n)-cos(b \xi_n)}    \Big], 
\end{equation}
where $\rho_0=\eta e^{i \xi_n}$, $\tau_n=\sqrt{|\sigma / \beta_1| \eta^2} \tau$, $\xi_n=|\sigma| \eta^2 \xi$;
$\eta$ is the amplitude background, $\Omega$ is the modulation frequency, $a=1/2(1-\Omega^2/4)$,
where $0 < a < 1/2 $ determines the frequency that experience gain, and $b=[8a (1-2a)]^{1/2}$ 
determines the instability growth. 

%
\begin{figure}[h!]
\begin{center}
\includegraphics[width=7.8cm]{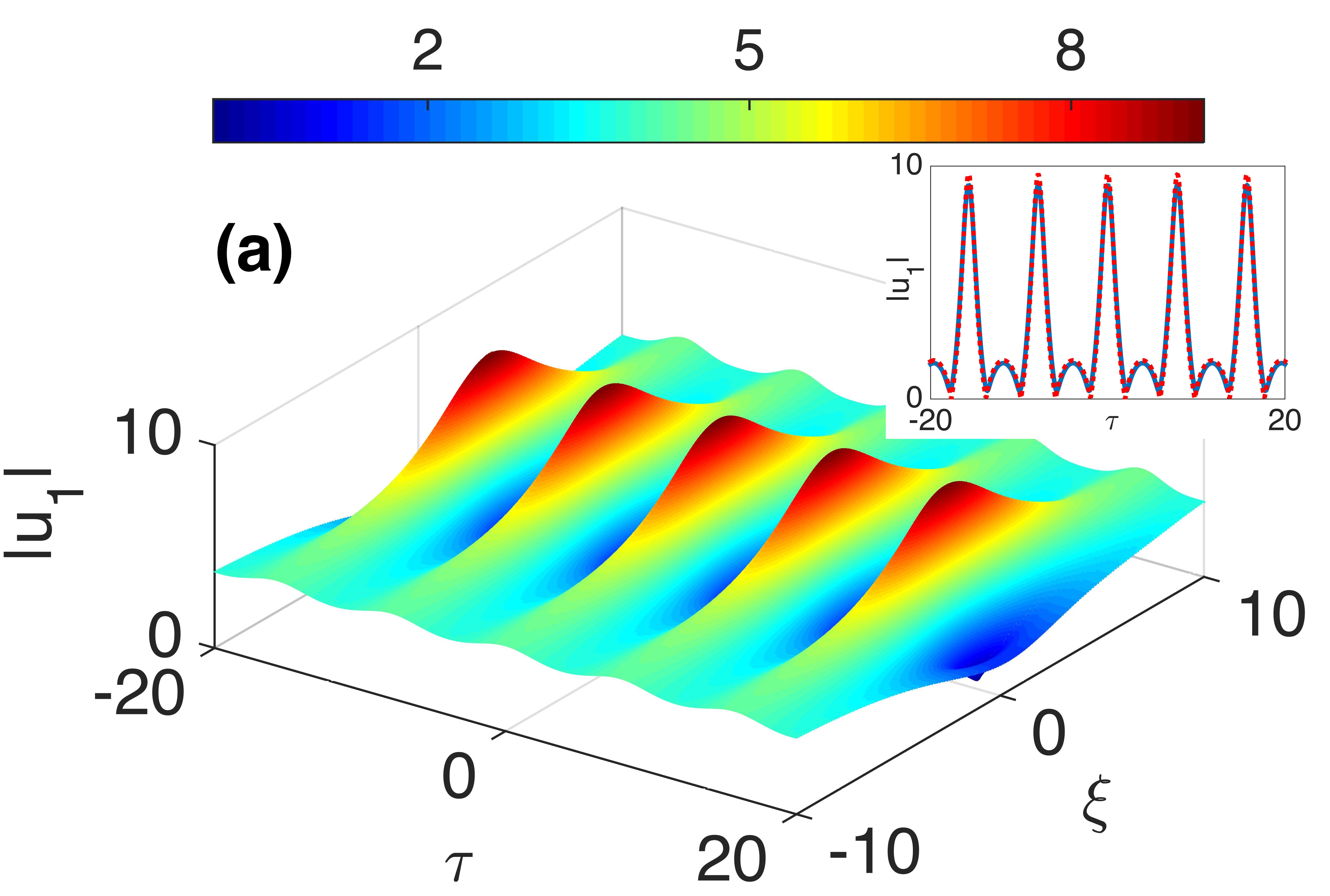}
\includegraphics[width=7.8cm]{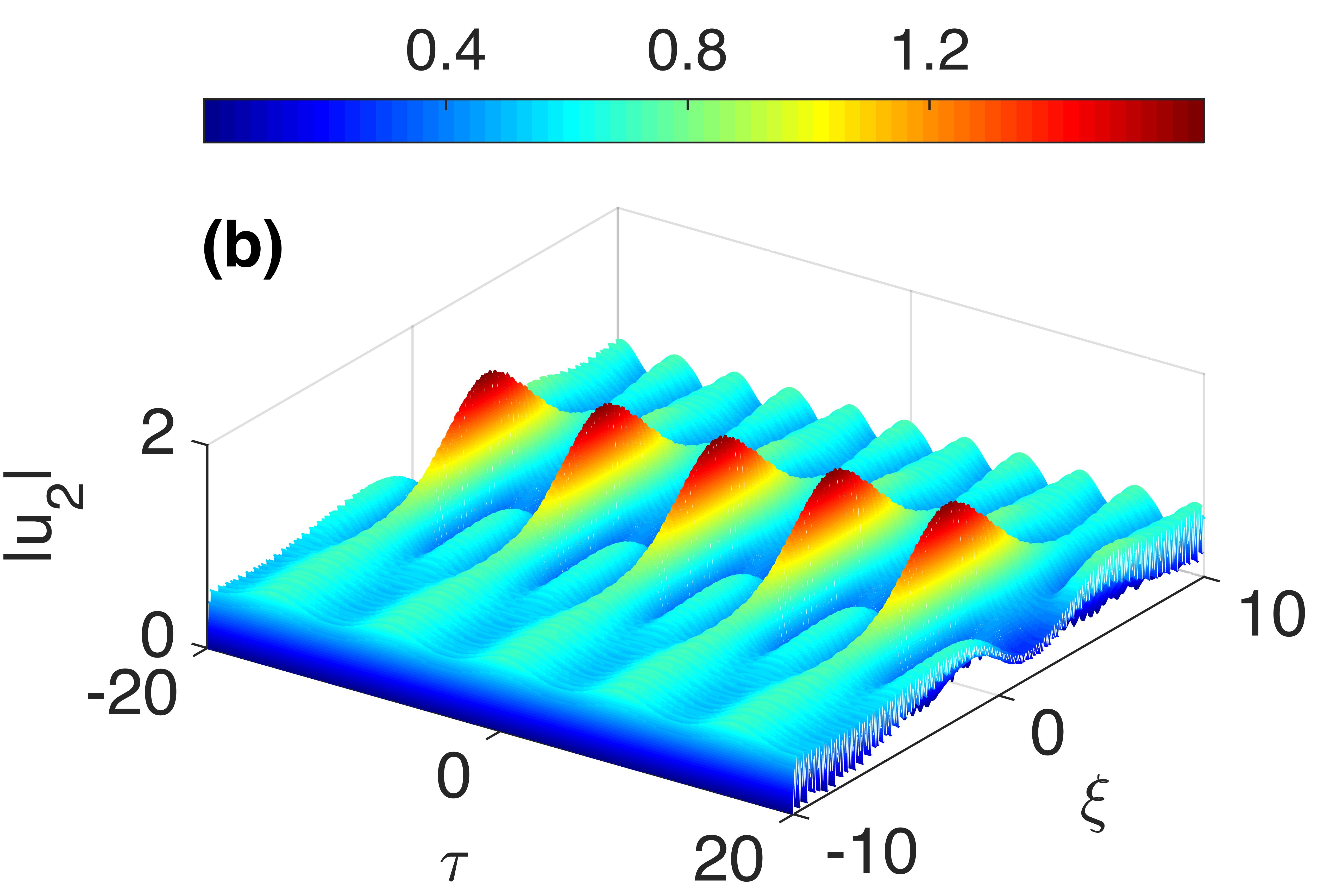}
\caption{(a), (b) Spatiotemporal numerical envelopes evolution, $|u_1|$  at FF and $|u_2|$ at SH, 
of a typical quadratic cascading Akhmediev breather solitary wave, in the $(\tau, \xi)$ plane. 
The inset reports the comparison between the numerical (blue line) and the analytical (dot red line)
envelope profile of the Akhmediev breather, at $\xi=0$.
Here $\beta_1=-1, \beta_2=-1, v=0, \delta k=50$; 
 $\eta=4, \Omega=\sqrt{2} (a=1/4, b=1)$.} \label{fig1}
  \end{center}
\end{figure}

We verified numerically the theorethical prediction of the FF Akhmediev breather dynamics $u_1=\rho$
of the Eqs. (\ref{shg}), where $\rho$ is the NLSE Akhmediev breather
soliton solution (\ref{AB}), with only the FF wave at the input of the quadratic medium.
Thus, in our numerics, the input solitary wave envelopes $u_{1,2}$  at $\xi=\xi_0$ are given by the expression
$u_1(\tau,\xi_0)=\rho(\tau,\xi_0)$, $u_2(\tau,\xi_0)=0$.

Figure \ref{fig1}(a), (b) report the spatiotemporal numerical envelopes evolution 
$|u_1|$  at FF, and $|u_2|$ at SH of a  typical quadratic cascading Akhmediev breather solitary wave,
in the $(\tau, \xi)$ plane. 
The numerical evolution of the FF Akhmediev breather in Fig. \ref{fig1}(a) maps very well on the 
evolution predicted by the soliton solution  (\ref{AB}) of  the NLSE Eq. (\ref{CLL}). 
The inset  in Fig. \ref{fig1}(a) reports the comparison between the numerical (blue line) and the analytical (dot red line)
envelope profile of the Akhmediev breather at his maximal amplitude, at $\xi=0$.
In this particular case, we estimate the relative error between the numerical FF wave of the Eqs. (\ref{shg}) 
and the analytical formula (\ref{AB}) to be around $4\%$, because of the FF depletion due to SH wave generation.

Very interestingly, the generated SH wave in Fig. \ref{fig1}(b), apart from low amplitude dispersive waves, 
is of breather nature and strictly locked with the FF component, indicating the existence of genuine quadratic Akhmediev solitary breathers.


%

Next, we proceeded to demonstrate the existence of Peregrine solitary waves of the SHG Eqs. (\ref{shg}),
again predicted theoretically through the mapping with the NLSE Eq. (\ref{CLL}).

A rational form of the  Peregrine soliton solutions of the Eq. (\ref{CLL}), under the condition $\beta_1 \sigma<0$,
can be expressed as \cite{pe83,ki10}:
\begin{equation}\label{PS}
\rho(\tau,\xi)=\rho_0 \Big[ 1+ \frac{4 (1+2 i \xi_n) }{1+ 4\tau_n^2 +4 \xi_n^2}    \Big],
\end{equation}
where $\rho_0=\eta e^{i \xi_n}$, $\tau_n=\sqrt{|\sigma / \beta_1| \eta^2} \tau$, $\xi_n=|\sigma| \eta^2 \xi$,
$\eta$ is the amplitude background. The amplitude $|\rho(\tau,\xi)|$ is peaked at $\xi=0$ with the maximum 
value $3|\eta|$ at $\tau=0$. Note that the rational solution (\ref{PS}) can be derived 
from (\ref{AB}), in the asymptotic case $a \rightarrow 1/2$.

\begin{figure}[h!]
\begin{center}
\includegraphics[width=7.8cm]{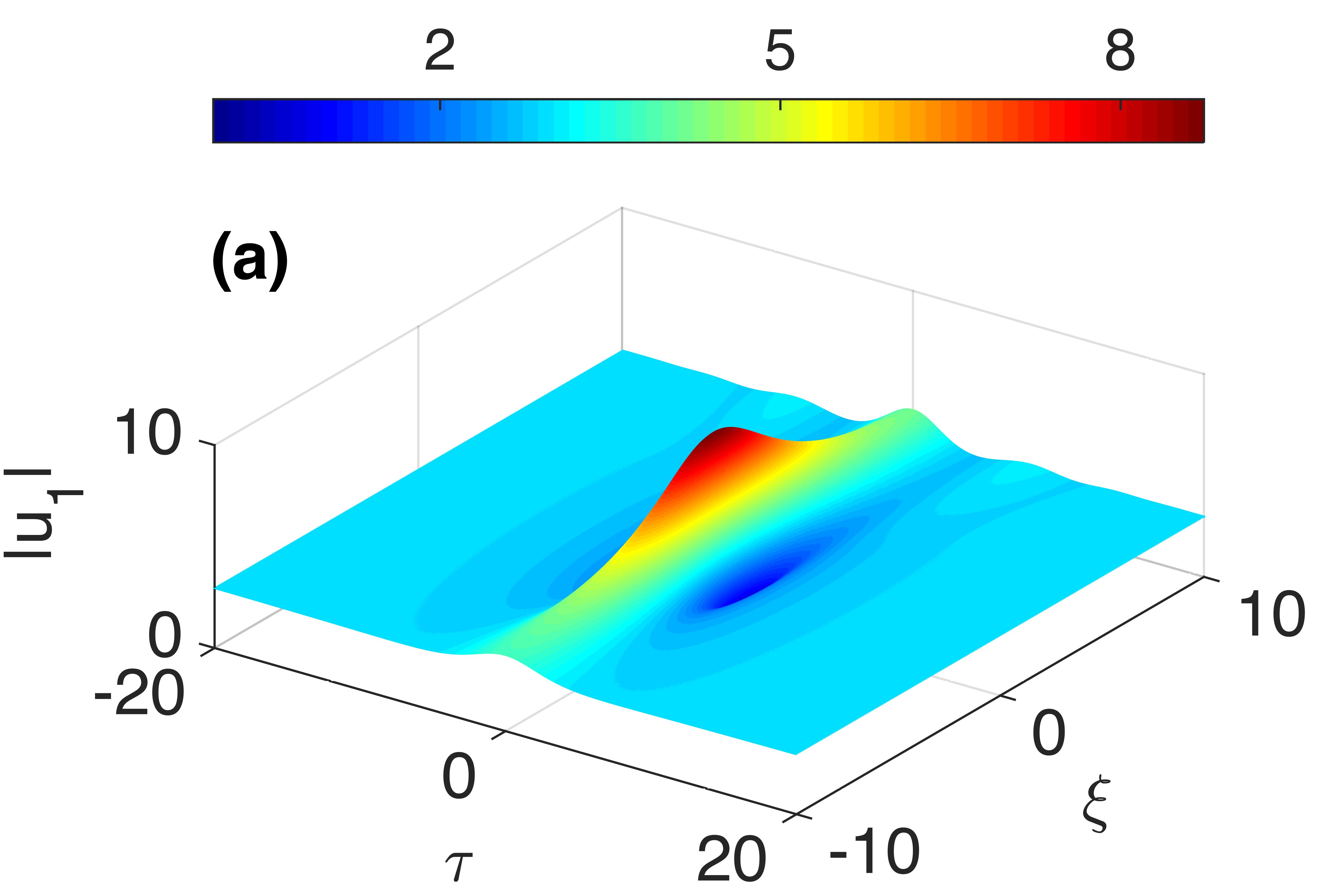}
\includegraphics[width=7.8cm]{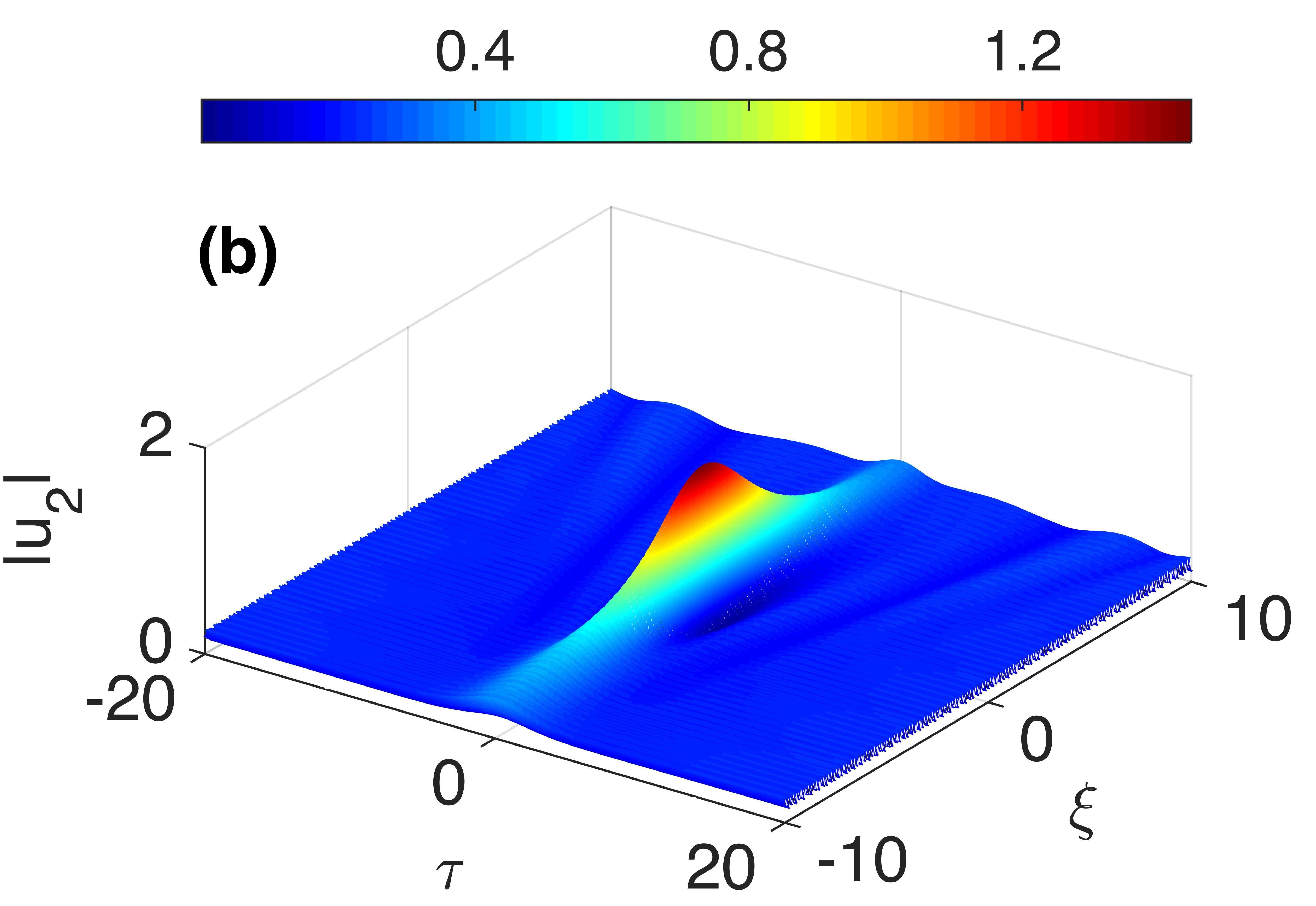}
\caption{
(a), (b) Spatiotemporal numerical envelopes evolution, $|u_1|$  at FF and $|u_2|$ at SH, 
of a typical quadratic cascading Peregrine solitary wave, in the $(\tau, \xi)$ plane. Here $\beta_1=1, \beta_2=1, v=0, \delta k=-50$ and 
 $\eta=2.5$.} \label{fig2}
  \end{center}
\end{figure}

In the case at hand, we assumed that the FF wave propagates tightly locked with the SH wave, and
we tested the theoretical prediction of a genuine quadratic Peregrine solitary dynamics $u_1=\rho$ at the FF, and 
$u_2=\rho^2e^{i \delta k \xi}/ \delta k$ at the SH, where $\rho$ is the NLSE Peregrine soliton solution (\ref{PS}).
Thus, in our numerics, the input solitary wave envelopes $u_{1,2}$  at $\xi=\xi_0$ are given by the expression
$u_1(\tau,\xi_0)=\rho(\tau,\xi_0)$, $u_2(\tau,\xi_0)=\rho^2(\tau,\xi_0) e^{i \delta k \xi_0}/\delta k$.

Figure \ref{fig2}(a), (b) report the spatiotemporal numerical envelopes evolution 
$|u_1|$  at FF, and $|u_2|$ at SH of a  typical quadratic cascading Peregrine solitary wave,
in the $(\tau, \xi)$ plane. The numerical evolution of the quadratic Peregrine components at 
the FF and SH in Fig. \ref{fig2} is very well predicted by the theoretical model.
We estimate the error between the numerical FF and SH wave of the Eqs. (\ref{shg}) 
and the analytical predictions to be lower than $1\%$.

MI in dispersive media with quadratic nonlinearity involves the spontaneous generation
of sideband pairs around both the FF and SH frequencies \cite{tri95}. 
MI may occur in the anomalous 
(i.e, $\beta_{1,2}<0$) and in the normal (i.e, $\beta_{1,2}>0$)
dispersion regimes; MI occurs also if the FF and the SH pulses are in different
regimes of dispersion (i.e., $\beta_1 \beta_2 <0$).
In the initial evolution of MI, the spectral sidebands associated with the instability experience exponential amplification at the expense
of the pumps, but the subsequent dynamics are more complex and display cyclic energy exchange between spectral modes \cite{tri97,shik00}.
In particular, MI seeded from noise results in a series of high-contrast peaks of different intensity.

\begin{figure}[h!]
\begin{center}
\includegraphics[width=7.8cm]{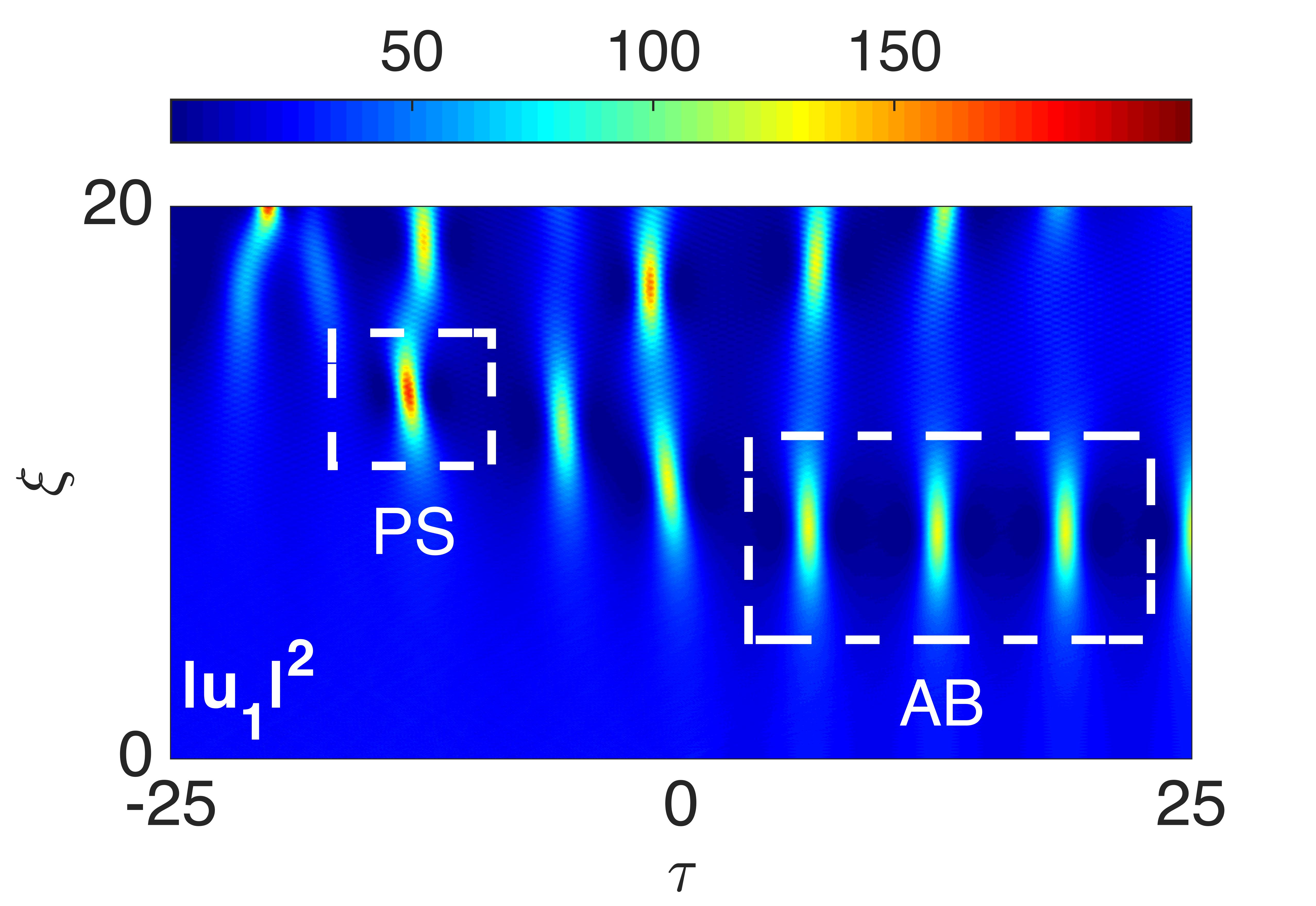}
\includegraphics[width=7.8cm]{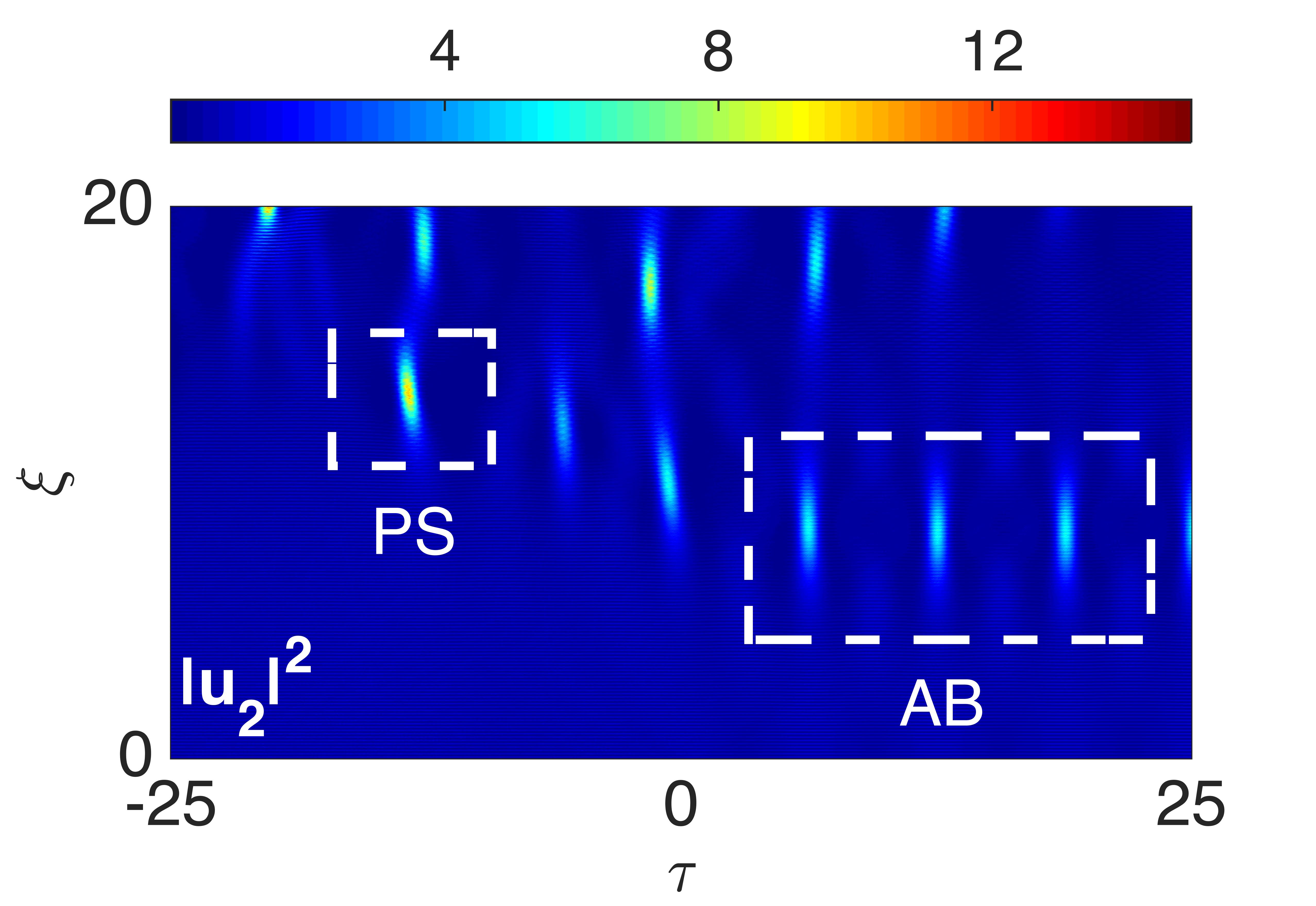}
\caption{
Density map  of the numerical temporal evolutions of FF (a) and SH (b) waves, in the $(\tau, \xi)$ plane. 
Here $\beta_1=-1$, $\beta_2=-1$, $v=0$, $\delta k=-50$. At the input, $u_1(\tau, 0)=u_ {10}+w$,
with $u_ {10}=5$, $w$ one-photon-per-mode noise, $u_2(\tau, 0)=0$. The label PS indicates Peregrine solitary
structures, the label AB indicates Akhmediev breathers.} \label{fig3}
 \end{center}
\end{figure}

Figure \ref{fig3}(a), (b) plot a density map of the numerical temporal evolutions of FF and SH waves, respectively, 
of a FF chaotic field triggered by one-photon-per-mode noise superimposed on a continuous-wave
background at the input of the quadratic medium. In our numerics, the input wave envelopes $u_{1,2}$  at $\xi_0$ 
are given by the expression $u_1(\tau,\xi_0)=u_ {10}+w$, $u_2(\tau,\xi_0)=0$, where $u_ {10}$
represent the continuous wave amplitude, $w$ the noise.
Figure \ref{fig3}(a), (b) report the emergence of an irregular series of temporal peaks,
triggered by MI, around $\xi=5$. Examining particular features of the evolution map (i.e., the periodicity
in $\tau$, $\tau-\xi$ localizations, the peak intensity values) reveals
clear signatures of quadratic Akhmediev solitary breathers and Peregrine solitary waves. 
Of course, observing ideal breathers or Peregrine structures is not expected 
given the random initial conditions, but it is remarkable how the quadratic breathers and Peregrine 
solitary solutions can be mapped closely to the noise-generated structures.
These results confirm that quadratic Akhmediev breathers and Peregrine waves can provide insights
into structures emerging from noise-seeded MI in SHG precesses.

Let us briefly discuss possible experimental settings in nonlinear optics for the observation
of quadratic extreme solitary wave dynamics. As to the temporal pulse dynamics,
one may consider optical temporal propagation in a bulk BBO, KTP or LiNbO$_3$ crystal in the regime of high phase mismatch, 
where quadratic solitary waves, temporal compression, spectral shift controls and steepening effects have been demonstrated 
(e.g., see the experimental setups of Refs. \cite{ditra98,moses06,baro06,bache07}). 
As to spatial dynamics, one may consider the spatial propagation in a bulk or in a planar
quadratic LiNbO$_3$ crystal, where spatial quadratic solitary waves, bouncing at
nonlinear interfaces, and solitary interactions have already been demonstrated
(e.g., see the experimental setups of Refs. \cite{schiek96,baro04}).

In conclusion, we investigated the formation of optical localized nonlinear structures, evolving upon 
a non-zero background plane wave, in  dispersive quadratic media.
We have demonstrated analytically and numerically the existence of quadratic Akhmediev 
breathers and Peregrine solitary waves, in the regime of cascading second-harmonic generation. 
On one hand, this finding opens a novel path for the prediction, excitation and control of extreme 
rogue waves in quadratic media; on the other hand, this result provides novel insights in MI dynamics. 

{\bf Funding} The present research was partially supported by the Italian
Ministry of University and Research (MIUR) (2012BFNWZ2).

\end{document}